\documentclass[lettersize,journal]{IEEEtran}
\usepackage{amsmath,amsfonts}
\usepackage{algorithmic}
\usepackage{array}

\usepackage{textcomp}
\usepackage{stfloats}
\usepackage{url}
\usepackage{booktabs}
\usepackage{verbatim}
\usepackage{graphicx}
\def\BibTeX{{\rm B\kern-.05em{\sc i\kern-.025em b}\kern-.08em
    T\kern-.1667em\lower.7ex\hbox{E}\kern-.125emX}}
\usepackage{balance}
\usepackage{booktabs}
\usepackage[numbers]{natbib}
\usepackage{subcaption}
\usepackage{float}
\usepackage{capt-of}

\usepackage[table]{xcolor}
\usepackage{booktabs}
\usepackage{multirow}

\definecolor{headerOrange}{RGB}{248, 203, 173}
\definecolor{headerGreen}{RGB}{198, 239, 206}
\definecolor{rowBlue}{RGB}{189, 215, 238}

\begin{document}
\title{Neural Signatures of Programming Expertise: Classifying Programmer Skill Levels Using EEG Data}

\author{
    Maurice~Rekrut,
    Mahima~Mahabaleshwar~Acharya,
    Taisiia~Ulianova,
    Norman~Peitek,
    Annabelle~Bergum,
    Mariya~Toneva,
    Sven~Apel,
    and~Antonio~Krüger
    \thanks{M. Rekrut, M. M. Acharya, T. Ulianova and A. Krüger are with the Cognitive Assistants Department, German Research Center for Artificial Intelligence (DFKI), Saarbrücken, Germany (e-mail: Maurice.Rekrut@dfki.de).}%
    \thanks{N. Peitek, A. Bergum and S. Apel are with Saarland University, Saarland Informatics Campus, Saarbrücken, Germany.}%
    \thanks{M. Toneva is with the Max Planck Institute for Software Systems, Saarbrücken, Germany.}
}

\maketitle

\begin{abstract}
Accurately assessing a programmer’s skill level is critical for hiring, team composition, and performance evaluation in the software industry. Conventional methods, such as coding tests or interviews, often fail to capture the full spectrum of cognitive abilities underlying programming expertise. This study explores using electroencephalography (EEG) and machine learning to investigate neural correlates of programming skill. We analyzed an existing EEG dataset recorded during code comprehension from 37 programmers with 1 to 30 years of experience (8.1 $\pm$ 6.3 years) to examine relationships between neural activity and expertise. Additionally, we conducted classification experiments using Random Forest classifiers with diverse features for binary (experts vs. novices) and multi-class (experts, intermediates, novices) setups. We identified EEG features and brain regions associated with programming expertise. Specifically, EEG entropy showed the strongest correlation with skill level. Furthermore, experts' brains were characterized by highly localized centro-frontal activation, whereas frontal activation in other groups was part of a more distributed network. Regarding classification, our setup achieved an average accuracy of 91.83\% (binary) and 78.15\% (multi-class) in stratified 10-fold cross-validation, while leave-one-subject-out validation achieved 85.00\% and 58.80\%, respectively. Individual frequency bands outperformed full-spectrum analyses, and both program comprehension and resting-state data yielded strong results. These findings demonstrate that EEG features effectively capture neural correlates across different skill levels and highlight the potential of neural data to complement traditional methods of skill assessment.
\end{abstract}


\section{Introduction}
Electroencephalography (EEG) has become a powerful tool for investigating cognitive processes such as attention, memory, and mental workload across diverse fields including psychology, education, and human–computer interaction. By capturing the brain’s electrical activity with high temporal precision, EEG enables researchers to study the neural correlates of expertise, decision-making, and learning in a non-invasive manner. In recent years, EEG-based methods have been increasingly employed to assess skill and expertise in domains, such as surgical performance~\cite{Shafiei:2023:Surgical}, gaming~\cite{anwar:2018:aGame}, and computer-aided design (CAD)~\cite{Baig:2019:Expertise, Baif:2020:Classification}. These studies suggest that EEG-derived features can serve as robust indicators of cognitive effort, task familiarity, and professional proficiency, offering a neurophysiological complement to behavioral assessments.
Expertise classification is a fundamental task in many applied contexts, where accurately determining skill level is crucial for optimizing performance, training, and recruitment. Traditional evaluation methods, such as self-assessments, interviews, and performance tests, often provide only partial insights into a person’s true cognitive potential. They are typically subjective, susceptible to bias, and limited to the evaluation of observable outcomes rather than the underlying cognitive mechanisms. EEG, by contrast, enables a more objective assessment of mental processes that unfold during task execution and can reveal the neural dynamics associated with expertise acquisition and performance optimization \cite{Crk:2014:Toward, Lee:2016:Comparing, Lee:2018:Mining, Shafiei:2023:Surgical}.

While EEG-based assessments have shown promise in several professional and creative domains, their application to programming expertise remains limited. Programming is a highly cognitive activity that engages a wide range of mental functions, such as working memory, sustained attention, abstraction, and problem-solving, that are reflected in neural activity patterns~\cite{Siegmund:2016:Program, Peitek:2022:Correlates, Duraisingam:2017:Cognitive, Goncales:2022:Empirical}. Yet, only a few studies have investigated how EEG features correlate with programming skill levels~\cite{Crk:2014:Toward, Crk:2016:Expertise, Lee:2018:Mining}, and those that do often rely on small datasets with a small range of in participant expertise, few EEG channels, or limited evaluation methods. Thus, the question whether neural correlates can distinguish between novice, intermediate, and expert programmers remains open.
Understanding the neuro-cognitive mechanisms underlying programming could provide valuable insights into how expertise develops and how individuals differ in problem-solving strategies. It could also inform the design of adaptive learning systems and training programs tailored to each programmer’s cognitive profile. By leveraging a large EEG dataset and machine learning, we aim to objectively model and predict programming skill levels, potentially complementing or enhancing existing assessment methods.

In the rapidly expanding software industry, accurately assessing a programmer’s skill level is essential for effective hiring, team composition, and performance evaluation. Conventional approaches such as portfolio reviews, coding tests, or interviews are inherently limited: they capture performance outcomes but not the mental processes that generate them \cite{Behroozi:2019:Hiring}. These methods are also prone to human biases and contextual influences. There is thus a need for objective and reliable measures that can provide deeper insights into programmers’ cognitive abilities.

EEG provides an opportunity to address this gap. By measuring brain activity during programming-related tasks, EEG can reveal moment-to-moment cognitive engagement, problem-solving strategies, and mental workload~\cite{Medeiros:2021:CanEEG}. When coupled with machine learning, EEG data can be used to automatically identify patterns of neural activity that correspond to different levels of programming expertise. Such models may eventually enable continuous, unbiased assessment of cognitive performance, supporting data-driven decisions in education, recruitment, and workforce development.
In addition to its practical relevance, studying EEG correlates of programming expertise also contributes to a more fundamental understanding of how cognitive skills manifest in neural dynamics. Specifically, it offers a window into how experience modifies the neural organization of problem-solving processes and how different brain regions interact during complex reasoning tasks.

This work investigates whether EEG data can be used to objectively distinguish between levels of programming expertise. Using a dataset of 37~participants with experience ranging from novices to experts (over 30 years of programming experience), recorded with a 64-channel EEG system, we analyze neural activity patterns during program-comprehension tasks. The paper makes the following key contributions:

\begin{itemize}
    \item Neural characterization of programming expertise: We identify EEG signal features that show significant correlations with programming skill level. The results reveal that entropy exhibits a strong correlation and best performance in classifying the skill level.
    
    \item Spatial and frequency-domain analysis: We provide a detailed analysis of EEG activity across frequency bands and scalp regions, demonstrating that individual frequency bands outperform full-spectrum features in discriminating expertise levels. Further, we show that centro-frontal regions are more active in expert programmers, while frontal activation is consistent across all groups.

    \item EEG-based skill classification framework: We implement and evaluate machine learning models for both binary (experts vs. novices) and multi-class (experts, intermediates, novices) classification. Using stratified 10-fold and leave-one-subject-out cross-validation, our models achieve accuracies up to 91.83\% and 85.00\%, respectively, highlighting the feasibility of EEG-based skill assessment.

    \item Implications for cognitive evaluation: We discuss the potential of EEG-driven metrics as complementary indicators of programming ability, with possible applications in personalized training, cognitive tutoring systems, and unbiased recruitment processes.
\end{itemize}

Together, these contributions demonstrate that EEG contains informative neural markers of programming expertise and that machine learning can effectively leverage these markers for skill classification. The findings open new directions for integrating neurophysiological data into software engineering research and practice, bridging cognitive neuroscience with human–computer interaction and computational learning.

\section{Background and Related Work}
Electroencephalography (EEG) has been widely used to investigate cognitive processes, mental workload, and task expertise across a broad range of domains, including education, psychology, human--computer interaction, medical surgery, and gaming. Numerous studies have demonstrated that EEG provides a sensitive means for capturing neural signatures related to attention, memory, and decision-making, thereby enabling the assessment of human performance during complex tasks \cite{Crk:2014:Toward, Lee:2016:Comparing, anwar:2018:aGame, Lee:2018:Mining, Baig:2019:Expertise, Baif:2020:Classification, Shafiei:2023:Surgical}. These works collectively show that EEG-based markers can be used to infer expertise levels in tasks ranging from surgical skill acquisition \cite{Shafiei:2023:Surgical} to gaming performance \cite{anwar:2018:aGame} and computer-aided design (CAD) workflows \cite{Baig:2019:Expertise, Baif:2020:Classification}. 

Beyond spectral differences, more recent studies have explored functional connectivity patterns and event-related potentials (ERPs) to quantify expertise. These methods have highlighted that experts often display more efficient neural processing, reflected through reduced cognitive effort or more focused activation patterns \cite{Baif:2020:Classification}. Together, these contributions suggest that EEG can serve as an objective tool for modeling cognitive proficiency and expertise across diverse real-world tasks.

\subsection{EEG for Assessment of Programming Expertise}

EEG has been used to study a wide range of phenomena in program comprehension, such as cognitive load~\cite{Medeiros:2021:CanEEG, Kosti:2018, Bergum:2025}, task difficulty~\cite{Duraisingam:2017:Cognitive, Kosti:2018, Palaniappan:2019}, confusion when understanding code~\cite{Yeh:2022, Bergum:2026}, machine learning classification of cognitive processes~\cite{Goncales:2020}, and concentration levels during pair programming~\cite{Thapaliya2020}.

The temporal resolution of EEG is particularly advantageous for expertise research, as it allows tracking rapid cognitive transitions that accompany problem solving and decision making. Furthermore, spectral features such as theta, alpha, and beta band activity have frequently been associated with task difficulty, learning progress, and domain mastery~\cite{Baig:2019:Expertise}. Several studies have leveraged EEG in such a way to obtain objective indicators of programming expertise. For example, \citet{Crk:2014:Toward} and \citet{Lee:2016:Comparing} showed that novices and experts exhibit different neural activation patterns when confronted with domain-relevant tasks, implying that EEG carries discriminative information about cognitive strategies and mental workload. Similarly, \citet{Crk:2016:Expertise} showed that EEG data could indicate the class level of students and their self-reported experience. These studies provide early evidence that the cognitive operations required for program comprehension, debugging, and logical reasoning elicit distinct EEG signatures depending on the programmer's experience level. 

\subsection{Neural Correlates of Programming Expertise}

Going beyond just EEG as a method, there is converging evidence that programming activates neural mechanisms related to working memory, sustained attention, and rule-based reasoning. Research in cognitive psychology and neuroscience has shown that program comprehension shares neural substrates with natural language processing and problem solving~\cite{Siegmund:2016:Program, Floyd:2017:Decoding, Peitek:2020, Castelhano:2019, Krueger:2020, Peitek:2022:Correlates}. EEG-based studies further support this perspective: For example, \citet{Duraisingam:2017:Cognitive} and \citet{Goncales:2022:Empirical} demonstrated that programmers show characteristic patterns of frontal and parietal activation during program-comprehension tasks, potentially reflecting cognitive load and domain-specific mental models. 

However, prior EEG studies aiming to identify neural correlates of expertise have been constrained by small sample sizes, limited EEG channel configurations, or narrow task designs. They focus on controlled stimuli or highly simplified programming tasks, limiting their ecological validity. Moreover, classification attempts have largely relied on basic machine learning techniques using relatively shallow feature sets, leading to modest predictive performance \cite{Crk:2014:Toward, Lee:2018:Mining}. Consequently, we still lack a systematic investigation of the neural correlates of programming skill across diverse task conditions and skill levels.

\subsection{Machine-Learning Classification of EEG-Data for Identifying Programmer Expertise}

The idea of using EEG for automatic classification of programmer skill level is still in its infancy. Early studies have attempted to distinguish novice and expert programmers using features derived from spectral power, ERPs, or simple temporal descriptors~\cite{Crk:2014:Toward, Lee:2018:Mining}. Although these works reported initial evidence that EEG carries discriminative information relevant to programming expertise, their classification accuracies and generalizability remain limited by restricted datasets and evaluation protocols.

Nonetheless, the overall direction of findings suggests that EEG may capture cognitive effort, strategy, and efficiency during programming tasks. For example, \citet{Lee:2018:Mining} found that experts exhibit reduced theta activity during program comprehension compared to novices, which the authors interpreted as more efficient cognitive control. Similarly, \citet{Goncales:2022:Empirical} observed differences in frontal theta and parietal alpha modulation across skill levels during debugging tasks. These results indicate that machine learning classifiers trained on EEG data could, in principle, identify neural markers that reliably distinguish between programmer expertise groups.

Despite promising developments, significant gaps remain. No existing study has systematically evaluated multiple EEG feature classes, task variations, and advanced classification models within a unified framework. Moreover, there is a shortage of larger datasets that include diverse programmer groups and realistic programming tasks. To advance the field, it is necessary to move beyond preliminary evidence toward a more comprehensive characterization of the neural foundations of programming expertise.

\subsection{Research Gap}

Overall, existing research demonstrates that EEG can reveal meaningful differences in cognitive processes underlying expertise across domains, including early explorations in programming. However, prior studies are fragmented, often limited in scope, and rarely integrate neural correlates, task-related features, and machine learning classification within a cohesive methodology.

The present study addresses this gap by developing a systematic framework for investigating neural signatures of programming skill. We aim to establish a more rigorous foundation for an EEG-based assessment of programming proficiency and to support future applications in education, recruitment, and adaptive learning technologies.

\section{Material and Methods}
This section describes the methodological framework used to analyse neural signatures of programming expertise and to evaluate the feasibility of classifying skill levels from EEG data. The workflow includes three main components: (i) correlation analysis between neural activity and programmer skill, (ii) spatial pattern extraction and (iii) supervised classification across binary and multi-class configurations of programmer skill score.

\subsection{Dataset}
For our study, we used the EEG data recorded by~\citet{Peitek:2022:Correlates}, which we describe in more detail next.

\subsubsection{Participants}
The dataset consists of EEG data of 37 participants with programming experience ranging from first year of programming at the university to 30 years of experience in industry. Expertise was quantified through a combination of self-reported years of experience, professional background, and validated coding performance metrics, based on a validated questionnaire for programming experience~\cite{Siegmund2014:Exp}. Table~\ref{tab:participant} provides an overview on participant's demographic data and programming expertise.

 \begin{table}[h!] 
    \centering
    \begin{tabular}{lr}
    \toprule
    \textbf{Participant Demographic/Experience Measure} & \textbf{No./Mean ± SD}\\
    \midrule
    Number of (Included) Participants & 37 \\
    Female & 5  \\
    Male & 31 \\
    Non-Binary  & 1\\
    Age (in Years) & 25.95 ± 6.76\\
    \specialrule{0.1pt}{1pt}{1pt}
    Undergraduate/graduate students & 27 of 37 \\
    . . . of which work (at least part time) & 14 of 27  \\
    Full-time professionals & 10 of 37 \\
    \specialrule{0.1pt}{1pt}{1pt}
    Years of Learning Programming & 7.93 ± 6.14 \\
    Years of Professional Programming & 3.55 ± 4.30 \\
    Years of Java Programming & 4.54 ± 4.31 \\
    Number of Known Programming Languages & 5.11 ± 2.02 \\
    \specialrule{0.1pt}{1pt}{1pt}
    Comparison to Peers$^\alpha$ & 3.67 ± 0.76 \\
    Comparison to 10-Year Professional$^\alpha$ & 2.25 ± 0.94 \\
    \specialrule{0.1pt}{1pt}{1pt}
    Hours per Week Spent in Software Engineering$^\beta$ & 24.76 ± 21.08 \\
    Hours per Week Spent Programming$^\beta$ & 10.78 ± 11.36\\
    \bottomrule
    \end{tabular}
    \caption{Overview of Peitek et al. \cite{Peitek:2022:Correlates} study's participant demographics and 8 out of 49 overall measures from their experience questionnaire. ${\alpha}$ denotes self-estimated measures on a 1-5 Likert scale~\cite{Robinson:2014:Likert}. ${\beta}$ denotes measures that were only collected from (at least part-time) professional programmers.}
    \label{tab:participant}
\end{table}

\subsubsection{Experimental Design and Protocol}

\begin{figure}
    \centering
    \includegraphics[width=1\linewidth]{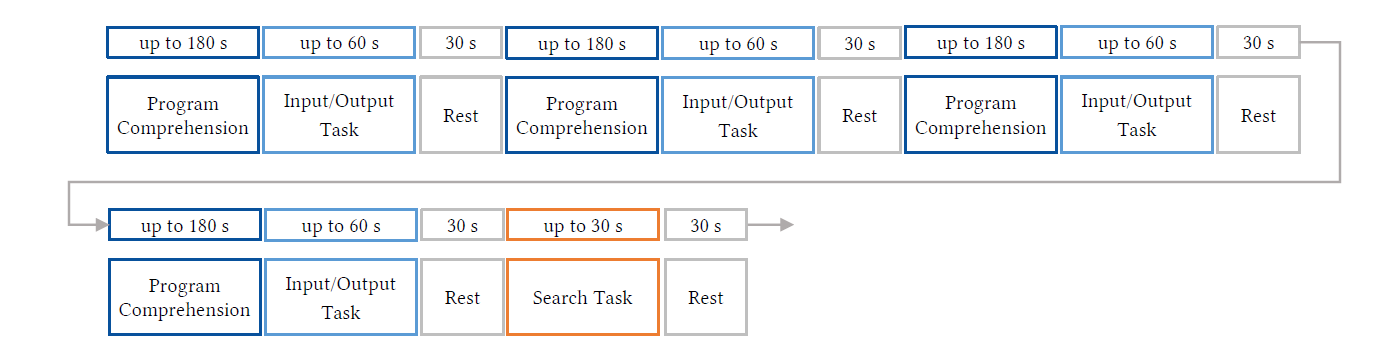}
    \caption{Visualization of the experimental design of Peitek 
 et al. \cite{Peitek:2022:Correlates} showing the first 4 comprehension tasks.}
    \label{fig:experimentalDesign}
\end{figure}

\citet{Peitek:2022:Correlates} conducted an EEG study with a within-subject experiment design involving 32 program-comprehension tasks, following prior work~\cite{Crk:2014:Toward, Lee:2018:Mining}. The tasks were presented in a pseudo-randomized order across three runs of approximately 20 minutes each, with optional 5-minute breaks to reduce fatigue. A search task requiring participants to count brackets was included as a baseline condition, following common practice in program comprehension research~\cite{Peitek:2021:Program}. The experiment followed a 4:1 ratio of comprehension to search tasks, and each task was separated by a 30-second resting phase during which participants fixated on a cross. We illustrate the experimental design in Figure~\ref{fig:experimentalDesign}. 

Peitek et al. conducted the study in a dim and distraction-free environment where participants sat in a comfortable chair to minimize muscle-related artifacts. EEG signals were acquired with a 64-channel LiveAmp~64 system at a 500~Hz sampling rate, using the international 10-20 electrode placement system~\cite{Jasper:1958:ReportOT}. References were placed at the mastoids, and electrode impedances were maintained below 5~k$\Omega$. The high-density montage enabled characterization of frontal, central, and parietal activity, which have been implicated in programming-related cognition \cite{Duraisingam:2017:Cognitive, Goncales:2022:Empirical}. The experimental protocol was implemented using PsychoPy~\cite{Peirce:2019:PsychoPy2}. After the session, a semi-structured interview captured participants’ subjective impressions of both the experiment and the presented code snippets.

The study received ethical approval from the ethical review board of the Faculty of Mathematics and Computer Science at Saarland University.
All participants provided informed consent and completed an experience questionnaire available in the replication package\footnote{\url{https://github.com/brains-on-code/NoviceVsExpert}}.

\subsubsection{Skill Level Assignment}

\begin{figure*}[t!]
    \centering
    \includegraphics[width=0.9\textwidth]{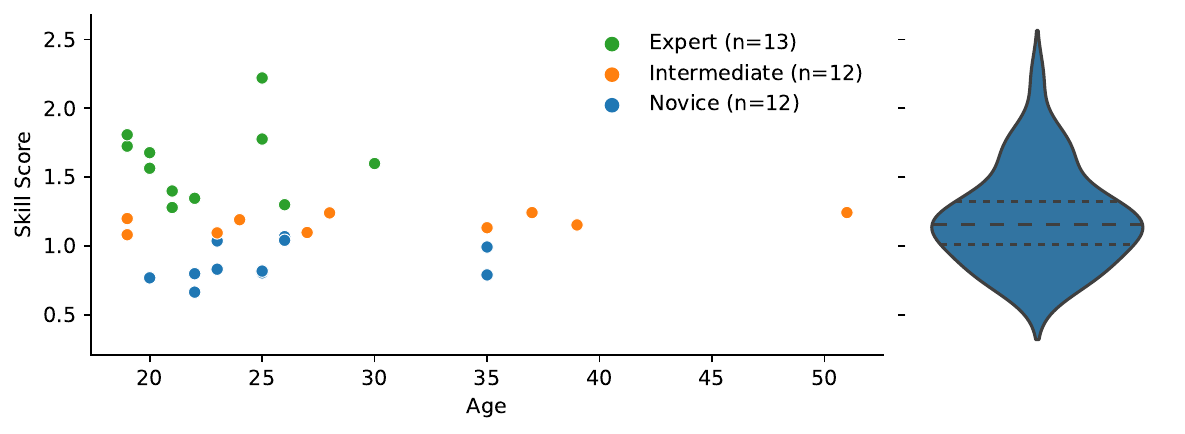}
    \caption{The skill score as distribution of correct answers per minute across participants, separated by the categorized expertise group and age of each participant.}
    \label{fig:skillScoreCombined}
\end{figure*}

We aimed at providing an objective measure for programmers skill level independent of the programming experience level (such as years of experience). To this end, we calculated a skill level based on the task performance during the program-comprehension tasks. Specifically, we categorized the responses of each participant for each task as either correct or incorrect. We then aggregated the total number of correct responses and the total time spent on the task for each participant. Our skill score is the number of correct answers per minute, following the approach of Peitek et al.~\cite{Peitek:2022:Correlates}. The distribution of the skill score indicates a substantial variability in skill level, which we visualize in Figure~\ref{fig:skillScoreCombined}.

Based on these skill scores, we assigned 37 participants to one of three skill levels. The lowest 33\% percentile of the distribution was labeled \textit{Novice} (n=12), the highest 33\% percentile \textit{Expert} (n=13), and the remaining participants \textit{Intermediate} (n=12).
To ensure that the observed EEG features are not confounded by age-related physiological differences, we assessed the age distribution across the three skill levels. The mean age was $25.58 \pm 4.75$ years for Novices, $30.20 \pm 10.19$ years for Intermediates, and $22.54 \pm 3.50$ years for Experts. A Kruskal-Wallis H-test revealed no significant difference in age between the groups ($H = 4.82, p = .089$). This suggests that the subsequent classification of skill levels is based on cognitive markers of expertise rather than age-correlated biological factors.

\subsection{Data Preprocessing}

In this section, we provide an overview of the procedures used for outlier removal, filtering, artifact rejection, epoching, and data assembly to prepare EEG data for analysis, generally following the pipeline established in Peitek et al.~\cite{Peitek:2022:Correlates}.

\subsubsection{Outlier Removal}
We removed outliers based on the original procedure described in Peitek et al.~\cite{Peitek:2022:Correlates}. Specifically, we examined the response times for each program-comprehension trial. We discarded the slowest 5\% (i.e., $> 2\,min~32\,s$) and fastest 5\% (i.e., $<11.s$). We only applied these thresholds to trials in which participants selected the ``Next'' option, as certain snippets can be answered quickly by experienced programmers. This ensured the removal of trials that likely reflected insufficient task engagement rather than genuine performance differences. 

Following this procedure, 9 trials from the lower 5\% and 55 from the upper 5\% were removed, resulting in 1072 remaining trials for further processing. One participant lost more than half of their data during this step and, therefore, was completely excluded from further analysis.

\subsubsection{Filtering}
We filtered the EEG signals to attenuate noise and physiological artifacts. In line with standard practice and previous work in EEG-based comprehension analysis~\cite{Rekrut:2020:Decoding}, we applied a Hamming-windowed finite impulse response (FIR) filter and a notch filter with a lower cutoff frequency of 49 Hz and an upper cutoff of 51\,Hz to remove power line noise. Additionally, we reduced baseline drift and high-frequency noise with a 4--200\,Hz bandpass filter. Finally, we applied common average referencing (CAR) by computing the average of the signal at all EEG electrodes and subtracting it from the EEG signal at every electrode for every time point. This approach reduces global noise and increases the spatial specificity of EEG by attenuating volume-conducted activity that is broadly distributed across the scalp~\cite{tsuchimoto2021use}.

\subsubsection{Rejecting Artifacts}
Because program comprehension requires a lot of eye movements~\cite{Sharif:2010, Busjahn2015, Peitek2020:Linearity}, we expected remaining residual eye and muscle artifacts after basic filtering. To address this, we performed a blind source separation (BSS)~\cite{Jutten:1991:BSS} using Independent Component Analysis (ICA) as implemented in the EEGLAB toolbox~\cite{Delorme:2004:EEGLAB}. ICA decomposes multichannel EEG into statistically independent components, enabling the removal of artifacts originating from eye movements, blinks, and muscular activity.
The components were automatically labeled using EEGLAB’s classification functions, and rejected based on confidence thresholds. Here, we removed eye- and muscle-related components using a probability threshold of $0.9$, while we discarded components without a clear class assignment at a threshold of $0.95$. This procedure ensured systematic and reproducible artifact removal without requiring a potentially biased manual inspection.

\subsubsection{Epoching}

\begin{figure}
    \centering
    \includegraphics[width=0.8\linewidth]{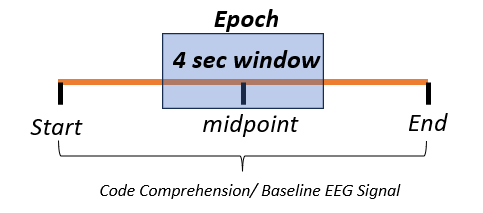}
    \caption{Visualization of the fixed 4-second epoching approach used in this study.}
    \label{fig:epoching}
\end{figure}

We epoched the EEG data to isolate neural activity associated with individual program-comprehension events. Using \texttt{mne.Epochs}\footnote{\url{https://mne.tools/stable/generated/mne.Epochs.html}}, we extracted 4-second fixed windows for each trial (cf.~Figure~\ref{fig:epoching}). We chose this duration based on the shortest task-completion time observed in the dataset, ensuring that all participants contributed equal-length data segments.
We centered epochs on the middle portion of each task interval, under the assumption that this segment best captures cognitive processing related to program comprehension, while minimizing onset- and offset-related artifacts. This approach improves comparability across participants and enhances the signal-to-noise ratio for subsequent analyses.

\subsection{Data Assembly}
\label{sec:data_assembly}

In the next step, we assembled the EEG data for analysis by systematically organizing the preprocessed signals into structured configurations that reflect different analytical perspectives. To this end, we followed two complementary approaches to feature extraction: task-based analysis and channel-based analysis. 

\subsubsection{Task-Based Analysis} 
In the task-based analysis, we focus on identifying how neural activity varies across the 32 program-comprehension tasks performed by each participant. For each task, EEG recordings from all 64 electrodes are segmented into the corresponding task window, followed by feature extraction per electrode. These features are then aggregated to form a single representative feature vector per task, capturing the participant’s neural response to the specific source-code snippet.
This approach enables investigation of task-dependent cognitive processing and supports comparison between different algorithms or levels of difficulty.

\subsubsection{Channel-Based Analysis}
A channel-based analysis emphasizes the spatial distribution of neural activity across cortical regions. Instead of grouping data by task, we aggregated all segments from all tasks per electrode, and extracted features across the resulting time series for each channel. This produces one feature vector per electrode, independent of specific task events.
This analysis highlights region-specific neural activity patterns and supports the exploration of spatial contributions to cognitive processing during program comprehension. 

\subsubsection{Program Comprehension and Baseline Analysis}
These two configurations were applied on the EEG data recorded during program comprehension.
We additionally conducted a baseline analysis to investigate whether programmer expertise can also be distinguished from resting-state neural activity. Between each program-comprehension trial, participants completed a 30-second eyes-open resting period. For each baseline segment, the fixed 4-second window centered within the resting-state interval was extracted, ensuring a consistent temporal snapshot across all participants and trials.

This setup resulted in four different data assembly strategies for further analysis, namely task- and channel-based assembly on the program comprehension and baseline data.
For each of the four configurations, we extracted different features as described in the following.

\subsection{Feature Extraction}
\label{sec:feature}

Feature extraction is a critical step in EEG analysis, providing quantitative descriptors of neural activity that enable subsequent classification and interpretation. In this work, we derive temporal, spectral, and statistical features to capture task-related variations in cognitive processing. These features form the basis for addressing the research questions regarding neural correlates of program-comprehension and individual differences in cognitive function.

\subsubsection{EEG Bandpower Features}
\label{sec:brainwaves}

EEG signals contain oscillatory activity across multiple frequency bands, each linked to distinct cognitive processes \cite{Lee:2016:Comparing}. In this study, we extracted bandpower features from the theta ($\theta$, 4--8 Hz), alpha ($\alpha$, 8--13 Hz), beta ($\beta$, 13--30 Hz), gamma ($\gamma$, 30--50 Hz), and broadband (4--50 Hz) ranges. 

We computed bandpower using Welch's method \cite{Welch:1967:FFT}, which estimates the power spectral density (PSD) by averaging periodograms computed over overlapping windowed segments. From the PSD, the relative power in each frequency band is calculated per channel, defined as the proportion of power in the respective band relative to the total spectral power \cite{Lee:2016:Comparing, Medeiros:2019:Software}. These normalized measures allow for robust comparison across participants and conditions.

\subsubsection{Statistical Features} 
\label{sec:stat_features}

For each frequency band’s relative power, we compute a set of descriptive statistical features using \texttt{numpy 1.26.1}\footnote{\url{https://numpy.org},~\cite{Harris2020}} and \texttt{scipy.stats 1.11.3}\footnote{\url{https://docs.scipy.org/doc/scipy/reference/stats.html},~\cite{SciPy2020}}. These features characterize the distributional properties of the spectral data:
    \textbf{Mean} (Average bandpower, representing overall activity level),
    \textbf{Median} (Central tendency robust to outliers),
    \textbf{Variance} (Degree of dispersion around the mean),
    \textbf{Standard deviation} ( Square root of variance, expressed in the data's original units),
    \textbf{Skewness} (Asymmetry of the distribution),
    \textbf{Kurtosis} (Degree of tail heaviness and presence of extreme values),
    \textbf{Zero-crossing rate} (Frequency of sign changes, reflecting fluctuation dynamics),
    \textbf{Entropy} (Measure of spectral unpredictability or complexity) 

In total, this results in 40 features per EEG channel (5 frequency bands × 8 statistical descriptors) for downstream analysis. We organized the extracted features in the before-mentioned configurations to support different analytical perspectives. This results in consistent feature vectors across all participants and analysis modes, enabling systematic evaluation of neural activity patterns.

\subsection{Data Analysis}
\label{sec:data_analysis}

We performed group-level analyses to address the core research objectives of this work to:
\begin{itemize}
    \item identify neural correlates between programmer expertise level and EEG features,
    \item perform topographical analyses to uncover activation patterns across skill groups,
    \item classify programmers skill level, independent of their behavior, only based on EEG data.
\end{itemize}

These analyses enabled identifying neural markers of programming proficiency, characterizing brain regions associated with program comprehension, and evaluating the viability of EEG-based skill classification. 

\subsubsection{Correlation Analysis} 
\label{sec:correlation_analysis}

A first objective of this study was to examine whether programmer expertise is reflected in systematic differences in neural features. To assess associations between skill level and EEG-derived measures, we computed Spearman's rank correlation coefficient $\rho$ between participants' level and the statistical features of bandpower extracted from each configuration using scipy.stats.spearmanr\footnote{\url{https://docs.scipy.org/doc/scipy/reference/generated/scipy.stats.spearmanr.html}} which evaluates a Spearman correlation coefficient with associated p-value. Spearman correlation is well suited for EEG data due to its robustness to non-normality and its ability to capture monotonic relations in nonlinear feature spaces~\cite{rousselet2012improving}. We treated skill levels as an ordinal variable (i.e., Novice, Intermediate, Expert), enabling direct assessment of whether increases in expertise correspond to consistent differences in neural activity across frequency bands and analysis configurations.

\subsubsection{Common Spatial Patterns} 
\label{CSP}

To identify spatial profiles of neural activity that distinguish program comprehension from resting state across different expertise levels, we applied Common Spatial Patterns (CSP) \cite{Koles:1991:quantitative, Blankertz:2008:Optimizing}. CSP provides spatial filters that maximize variance differences between two conditions, in this case, program comprehension versus baseline epochs, yielding interpretable topographies that reflect task-related engagement of cortical regions, calculated for each skill level. Following established practice in binary EEG paradigms \cite{Pfurtscheller:1999:EEG}, we retained four spatial components to capture discriminative patterns while avoiding overfitting. CSP thus serves as a complementary analysis to correlation-based feature inspection, enabling localization of expertise-dependent differences in task-related brain dynamics.

\subsubsection{Classification of Programmer Expertise} 
\label{Classification}

A central goal of this work was to evaluate whether EEG activity contains sufficient information to distinguish between programmer skill levels. We trained a Random Forrest (RF) classifier to predict either (i) two-level expertise (i.e., Novice vs. Expert) or (ii) three-level expertise (i.e., Novice, Intermediate, Expert). 
The choice of a Random Forest classifier is motivated by its proven robustness in handling high-dimensional and non-linear EEG features \cite{Goncales:2022:Empirical, Shafiei:2023:Surgical}. Previous empirical evaluations in the domain of program comprehension have identified RF as a top-performing model, often surpassing other classical machine learning algorithms in terms of accuracy and generalization \cite{Goncales:2022:Empirical}. Furthermore, the effectiveness of using non-deep learning classifiers for distinguishing between novice and expert cognitive states has been well-documented in both programming \cite{Lee:2016:Comparing} and other complex skill-based domains \cite{Shafiei:2023:Surgical}. This aligns with our objective to utilize an established, interpretable model that performs reliably on tabular neural features without the extensive data requirements of deep learning architectures.

To estimate generalization performance, we employed two cross-validation schemes targeting different aspects of robustness: stratified $K$-fold (SKF) for balanced within-dataset evaluation in our case implemented with 10 folds and leave-one-subject-out (LOSO) to assess performance on unseen individuals. 
These complementary validation strategies allow a systematic assessment of whether skill-related neural signatures generalize across participants.

We quantified model performance using standard metrics for multi-class EEG classification, including accuracy and F1-score. We performed all evaluations separately for each data assembly configuration (task-based, channel-based, on program comprehension and baseline data) to determine which neural representations provide the strongest basis for distinguishing programmer expertise.

\section{Results}
In this section, we present the results structured along the three core research objectives.

\subsection{Correlation of Programmer Expertise and EEG Features}
\begin{figure*}[t!]
     \centering
     \begin{minipage}{0.49\textwidth}
         \centering
         \includegraphics[width=\textwidth]{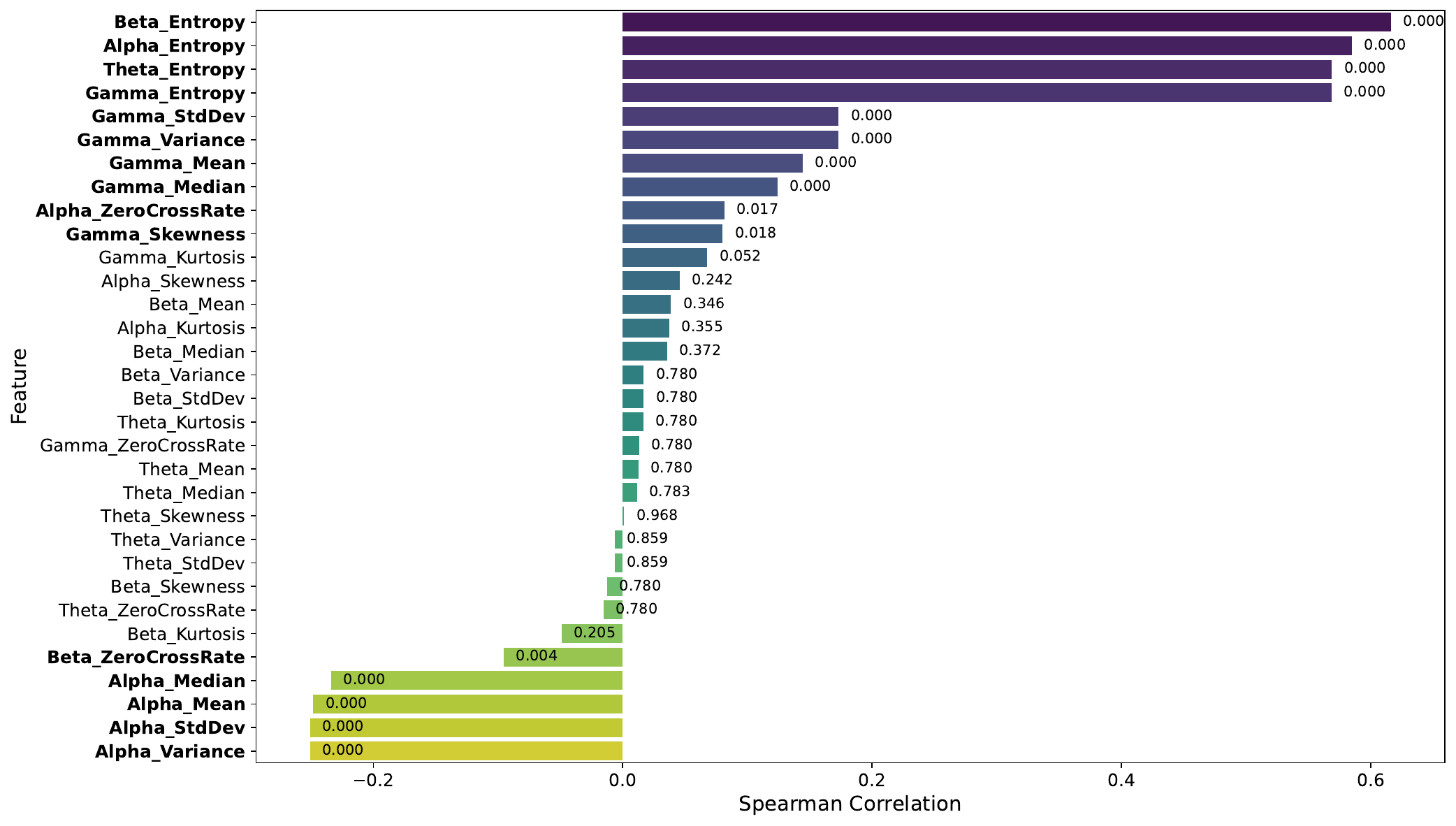}
         \caption*{Task-wise assembly baseline data.}
     \end{minipage}
     \hfill 
     \begin{minipage}{0.49\textwidth}
         \centering
         \includegraphics[width=\textwidth]{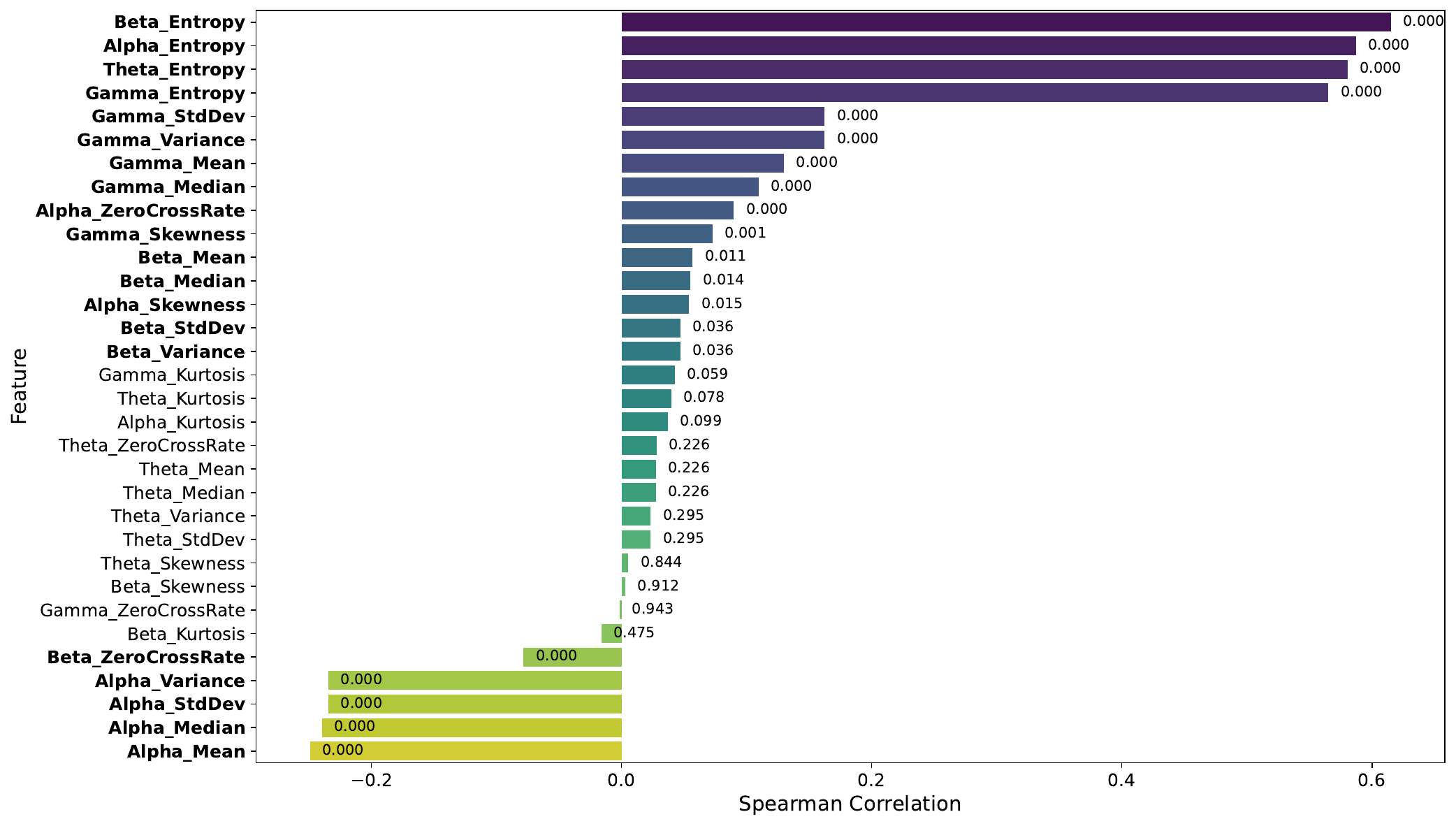}
         \caption*{Electrode-wise assembly baseline data.}
     \end{minipage}

     \vspace{1em} 

     \begin{minipage}{0.49\textwidth}
         \centering
         \includegraphics[width=\textwidth]{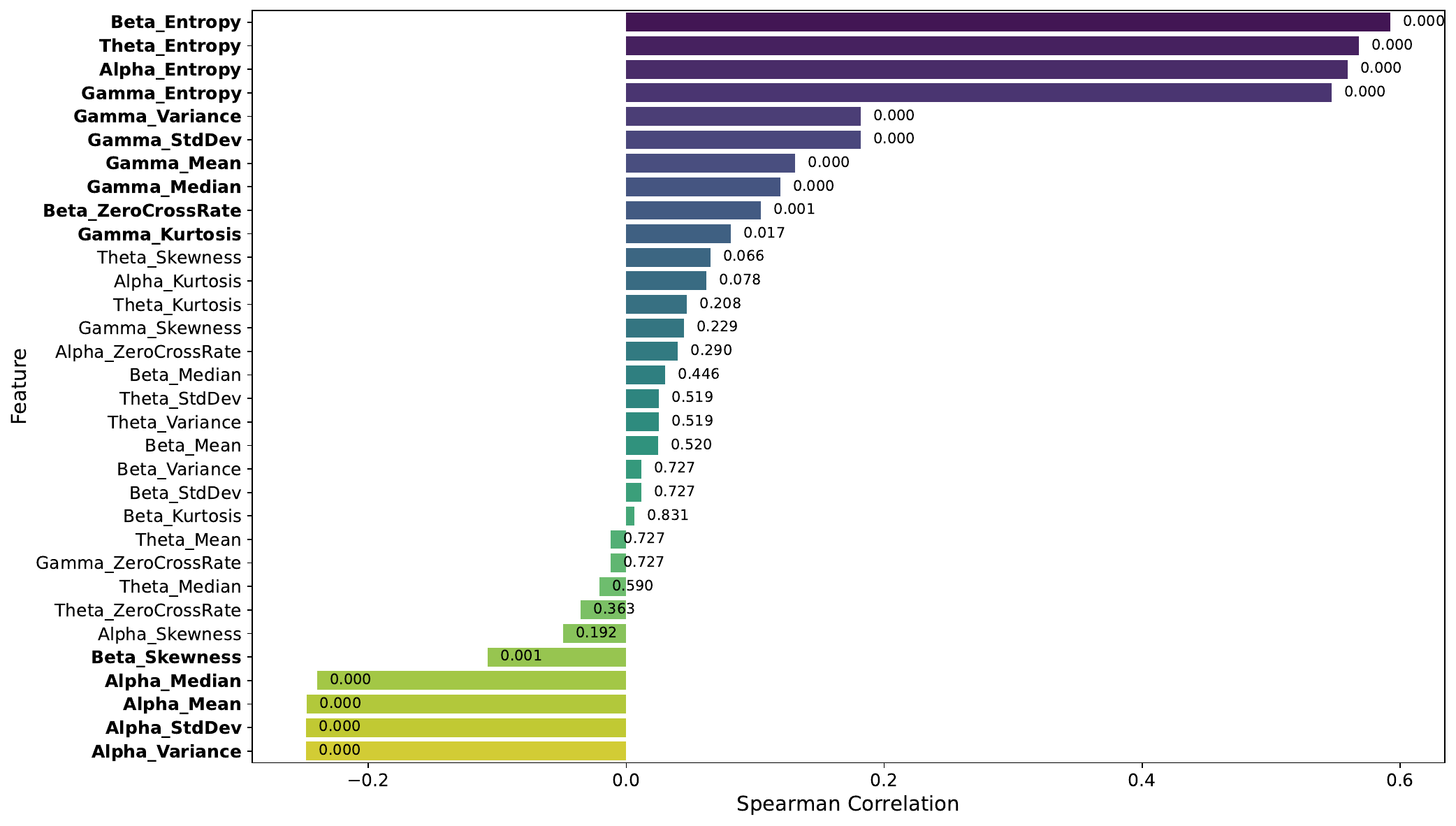}
         \caption*{Task-wise assembly program comprehension.}
     \end{minipage}
     \hfill
     \begin{minipage}{0.49\textwidth}
         \centering
         \includegraphics[width=\textwidth]{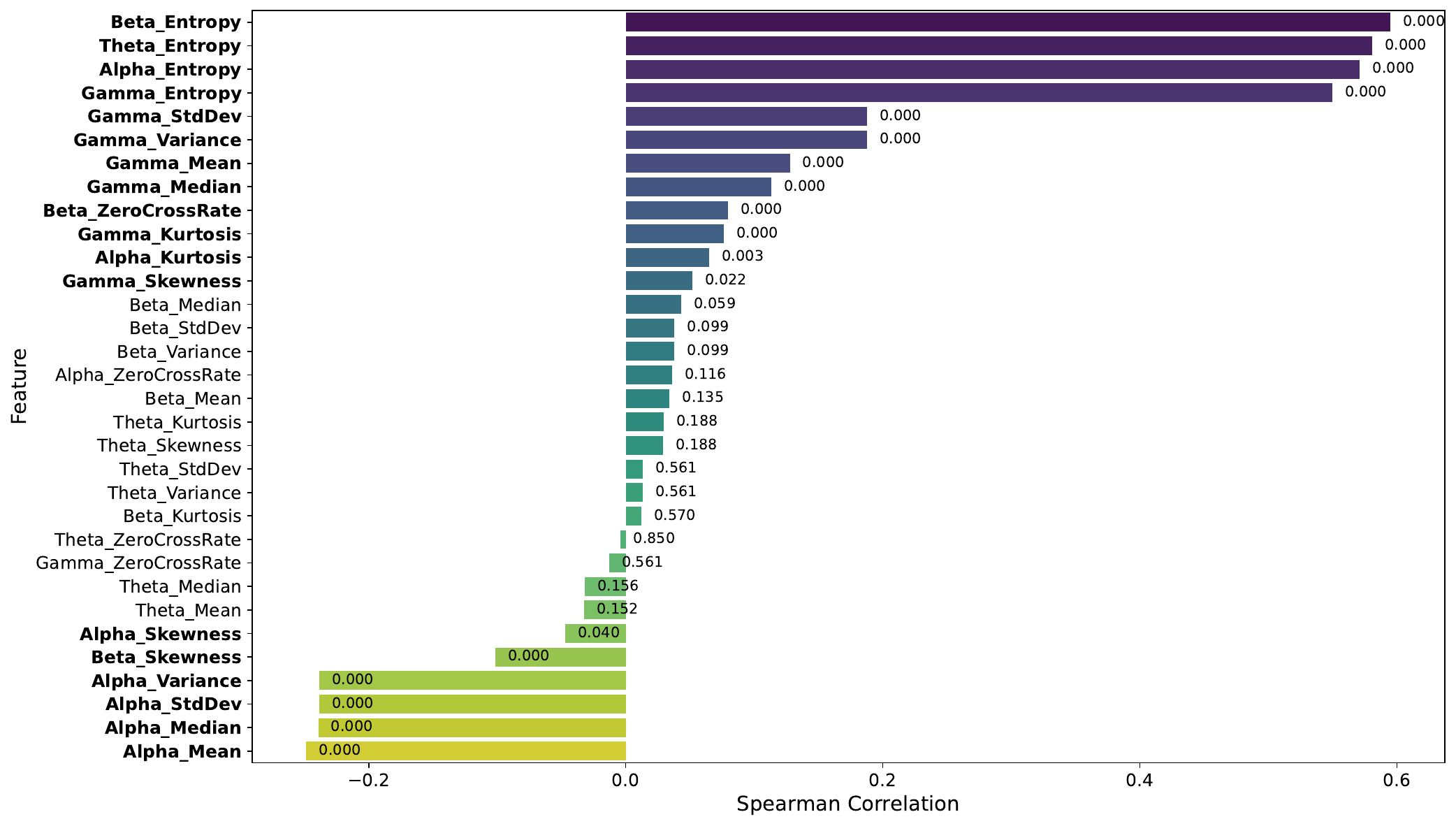}
         \caption*{Electrode-wise assembly program comprehension.}
     \end{minipage}
     
     \caption{Results of the Spearman Correlation analysis for baseline data (top row) and program comprehension data (bottom row) for the task- and electrode-wise data assembly conditions. Statistically significant features are highlighted in bold, the number behind each bar denotes the corresponding p-value.}
     \label{fig:correlation_results}
\end{figure*}

We performed a correlation analysis using Spearman Correlation on the extracted features as described in Section~\ref{sec:correlation_analysis} in four different conditions. Once on the baseline and once on the program-comprehension data, both for task- and electrode-specific arrangement. Figure~\ref{fig:correlation_results} shows the results of this correlation analysis, highlighting features that are statistically significant (p-value $<0.05$). The correlation analysis across all four experimental conditions, comprising both baseline and program-comprehension data, reveals a consistent relationship between specific EEG features and programming expertise. 

A primary finding is the robust performance of entropy features across the $\alpha, \beta, \theta,$ and $\gamma$ bands. In every scenario, entropy exhibited the highest positive correlation with skill level, maintaining a Spearman coefficient of $\rho \approx 0.6$. This suggests that signal complexity across the entire frequency spectrum serves as a reliable predictor of a programmer's experience. Furthermore, features derived from $\alpha$ and $\gamma$ waves emerged as the dominant predictors. This is particularly evident in the electrode-focused baseline data, where the number of significant features peaked at 21 out of 32.

In contrast, the analysis identifies a consistent set of features that fail to provide reliable indications of expertise. Statistical measures such as $\theta$ mean, $\theta$ median, and various Kurtosis values, specifically within the $\theta$ and $\beta$ bands, demonstrated no significant correlation across most trials. While the baseline data showed slight variances in significant $\beta$ features depending on the focusing method, the program-comprehension tasks confirmed that $\alpha$ and $\gamma$ attributes, such as standard deviation, variance, and mean, remain the most stable markers of skill level in programming contexts. Overall, the transition from algorithm-focused to electrode-focused pre-processing tended to slightly increase the density of significant correlations, especially within the $\beta$ band.

\subsection{Topographic Analysis of Programming Expertise}

\begin{figure*}[ht]
     \centering
     \begin{minipage}{0.9\textwidth}
         \centering
         \includegraphics[width=\textwidth]{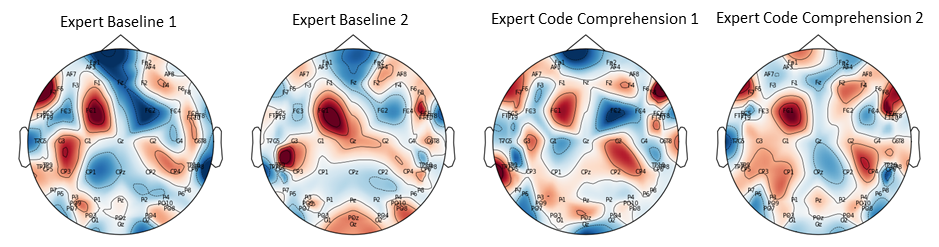}
         \caption*{Topoplots of expert programmers.}
         \label{fig:csp_results_expert}
     \end{minipage}
     
     \vspace{1.5em}

     \begin{minipage}{0.9\textwidth}
         \centering
         \includegraphics[width=\textwidth]{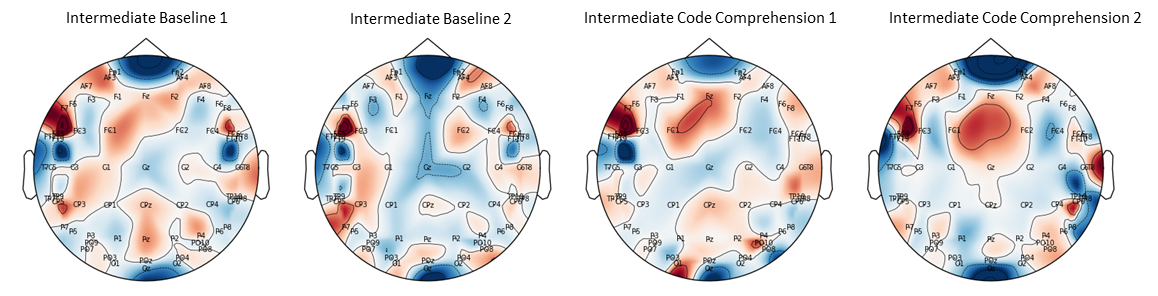}
         \caption*{Topoplots of intermediate programmes.}
         \label{fig:csp_results_intermediate}
     \end{minipage}

     \vspace{1.5em}

     \begin{minipage}{0.9\textwidth}
         \centering
         \includegraphics[width=\textwidth]{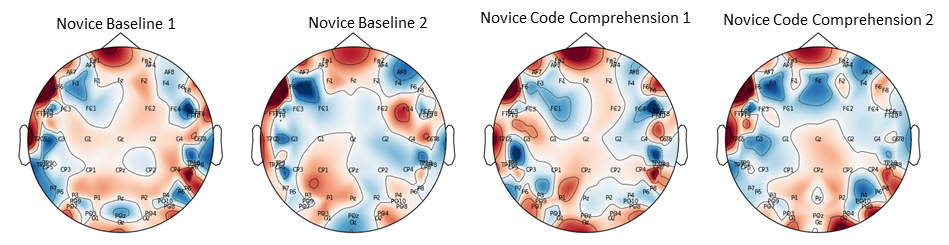}
         \caption*{Topoplots of novice programmers.}
         \label{fig:csp_results_novice}
     \end{minipage}
     
     \caption{Topoplots of CSP analysis for the three groups of programmers. First and second plot show baseline, third and fourth plot the program-comprehension components.}
     \label{fig:csp_results}
\end{figure*}

To address our second goal, we generated CSP topographical patterns using \texttt{mne 1.5.1}\footnote{Specifically, the \texttt{mne.viz.plot\_topomap}}. We illustrate the spatial filters in Figure~\ref{fig:csp_results}, where the color intensity represents the spatial filter weights. These weights highlight the regions of maximal variance that distinguish the baseline from the program-comprehension task. A comparative visual inspection reveals distinct strategies in neural processing across expertise levels. 

For experts, a remarkable stability is observed between the baseline (components 1--2) and the task-related patterns (components 3--4). The weights are consistently concentrated in the centro-frontal region, notably FC1. This suggests a highly efficient and automated mental set for program comprehension, where the transition from rest to task requires minimal topographical reconfiguration of neural resources.
In contrast, intermediate programmers show a more pronounced shift. While their baseline appears more diffuse, the task-related components exhibit a sharp, localized focus in the frontal area (channels F7, FC5, FC3). This indicates a clear transition into an active, concentrated cognitive state specifically for the task.
The novices display the most complex and widely distributed patterns. Unlike the localized activity of experts, novices exhibit high weights across frontal, temporal, and parietal regions (e.g., P1, T7, P8). The increased involvement of parietal areas suggests a higher reliance on visual scanning and basic syntax decoding. Furthermore, the strong activation in pre-frontal channels (Fp1, Fp2) likely reflects increased ocular activity, e.g. searching movements, potentially indicating residual EOG artifacts despite ICA pre-processing.

The CSP analysis identifies a clear progression, as expertise increases, neural involvement shifts from a distributed, vision-heavy parietal network (Novices) to a highly localized and efficient centro-frontal focus (Experts). These findings indicate that specific brain regions, particularly in the frontal cortex, serve as critical markers for programming expertise.

\subsection{Machine-Learning Classification of Programmer Expertise}

We evaluated the performance of the expertise classification using both binary (Novice vs. Expert) and multi-class (Novice, Intermediate, Expert) paradigms. We employed two validation strategies, namely stratified 10-fold cross-validation (SKF) and leave-one-subject-out (LOSO) cross-validation.

\begin{table*}[htbp]
    \centering
    \caption{Classification results for the different data assembly structures (task- and electrode-wise assembly) and evaluation conditions (SKF=stratified 10 fold; LOSO = leave one subject out) for program comprehension and baseline data.}
    \label{tab:classification_results}
    \small
    \renewcommand{\arraystretch}{1.2}
    \begin{tabular}{l|ccc|ccc}
        & \multicolumn{3}{c|}{\cellcolor{headerOrange} \textbf{Program Comprehension}} & \multicolumn{3}{c}{\cellcolor{headerGreen} \textbf{Baseline}} \\
        \cline{2-7}
        & \textbf{Task} & \textbf{Electrode} & \textbf{Average} & \textbf{Task} & \textbf{Electrode} & \textbf{Average} \\
        \hline
        \rowcolor{rowBlue} \multicolumn{7}{c}{\textbf{Binary Classification}} \\
        \hline
        \textbf{SKF} & 91.60 & 92.06 & 91.83 & 91.47 & 92.37 & 91.92 \\
        \textbf{LOSO} & 86.00 & 84.00 & 85.00 & 84.50 & 83.81 & 84.16 \\
        \textbf{Average} & 88.80 & 88.03 & 88.42 & 87.99 & 88.09 & 88.04 \\
        \hline
        \rowcolor{rowBlue} \multicolumn{7}{c}{\textbf{Multi-Class Classification}} \\
        \hline
        \textbf{SKF} & 76.77 & 79.52 & 78.15 & 80.04 & 82.39 & 79.37 \\
        \textbf{LOSO} & 59.86 & 57.73 & 58.80 & 61.48 & 60.14 & 60.81 \\
        \textbf{Average} & 68.32 & 68.63 & 68.47 & 70.76 & 71.27 & 71.01 \\
        \bottomrule
    \end{tabular}
\end{table*}

We show the results in Table~\ref{tab:classification_results}, which indicate that the binary classification achieved high accuracy, peaking at 92.06\% using the SKF method for program comprehension and 92.37\% for baseline data. Even under the more rigorous LOSO validation, which tests for generalizability across unseen subjects, a robust average accuracy of 85.00\% was maintained for program-comprehension tasks and 84.16\% for the baseline. 
In the multi-class classification, performance naturally decreased due to the higher complexity of the three-class problem. The SKF method reached an average accuracy of 78.15\%, while LOSO accuracy dropped to approximately 58.80\% for program comprehension and 79.37\% and 60.81\% for the baseline, respectively. 

Regarding the feature aggregation methods, the results show negligible variance between task-focused and electrode-focused assemblies. For instance, in the binary SKF classification of program-comprehension data, the electrode-focused approach (92.06\%) outperformed the task-focused approach (91.60\%) by only 0.46 percentage points. Across all experimental conditions and validation strategies, this marginal difference remained consistent, suggesting that both aggregation methods capture the underlying expertise-related neural patterns to a comparable degree.

The results for the baseline data were nearly identical to the program-comprehension data, indicating that the physiological markers of expertise are present even in resting states. But it is possible that the baseline not being long enough to ensure the users returning to a real rest-condition free of any program comprehension related brain activity. For example, \citet{Peitek:2018:fMRI} reported that participants reflect on a programming task during the subsequent rest condition. To investigate this issue in our data set, we conducted a temporal analysis across different positions of time windows which were taken from the start, the middle and the end of the baseline measurement. A key hypothesis was that the Start window of the baseline would yield higher classification accuracies due to ``neural resonance'' or lingering cognitive activity from the preceding task. Conversely, we expected a decline in discriminative features in the end window as the participants reached a deeper state of relaxation. Additionally, we applied temporal analysis with the three time windows on the program-comprehension data with the opposite hypothesis, that classification accuracy should improve towards the end of the task, due to participants being longer involved and establishing stronger activation patterns in brain regions related to program comprehension.

\begin{figure*}[t!]
     \centering  \includegraphics[width=0.9\textwidth]{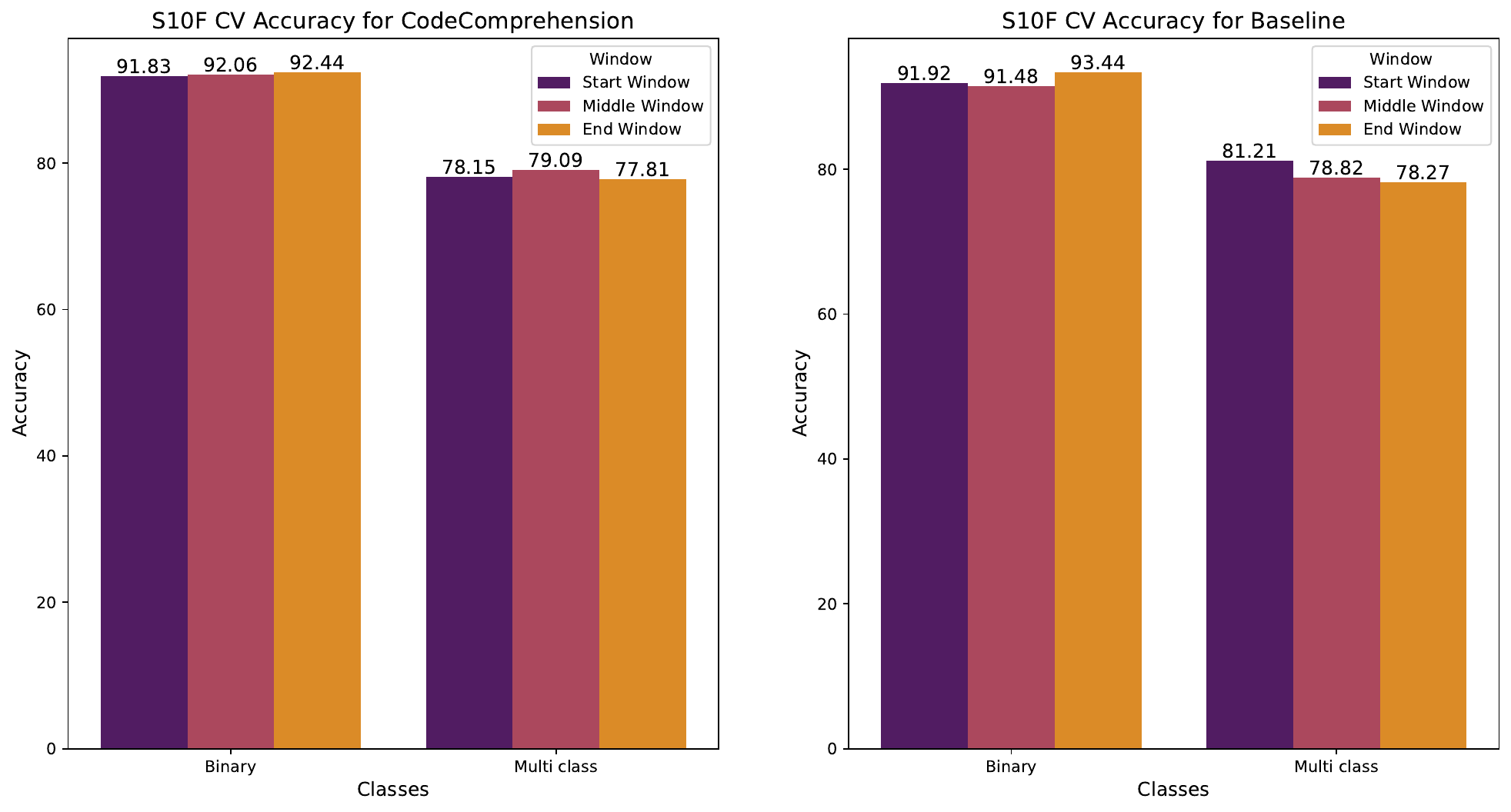}
     \caption{Comparison of different time window positions start, middle and end with stratified 10 fold evaluation for binary and multi-class classification. Program-comprehension data results are shown on the left, baseline data results on the right side.}    \label{fig:skf_win_results}
 \end{figure*}

 \begin{figure*}[t!]
     \centering \includegraphics[width=0.9\textwidth]{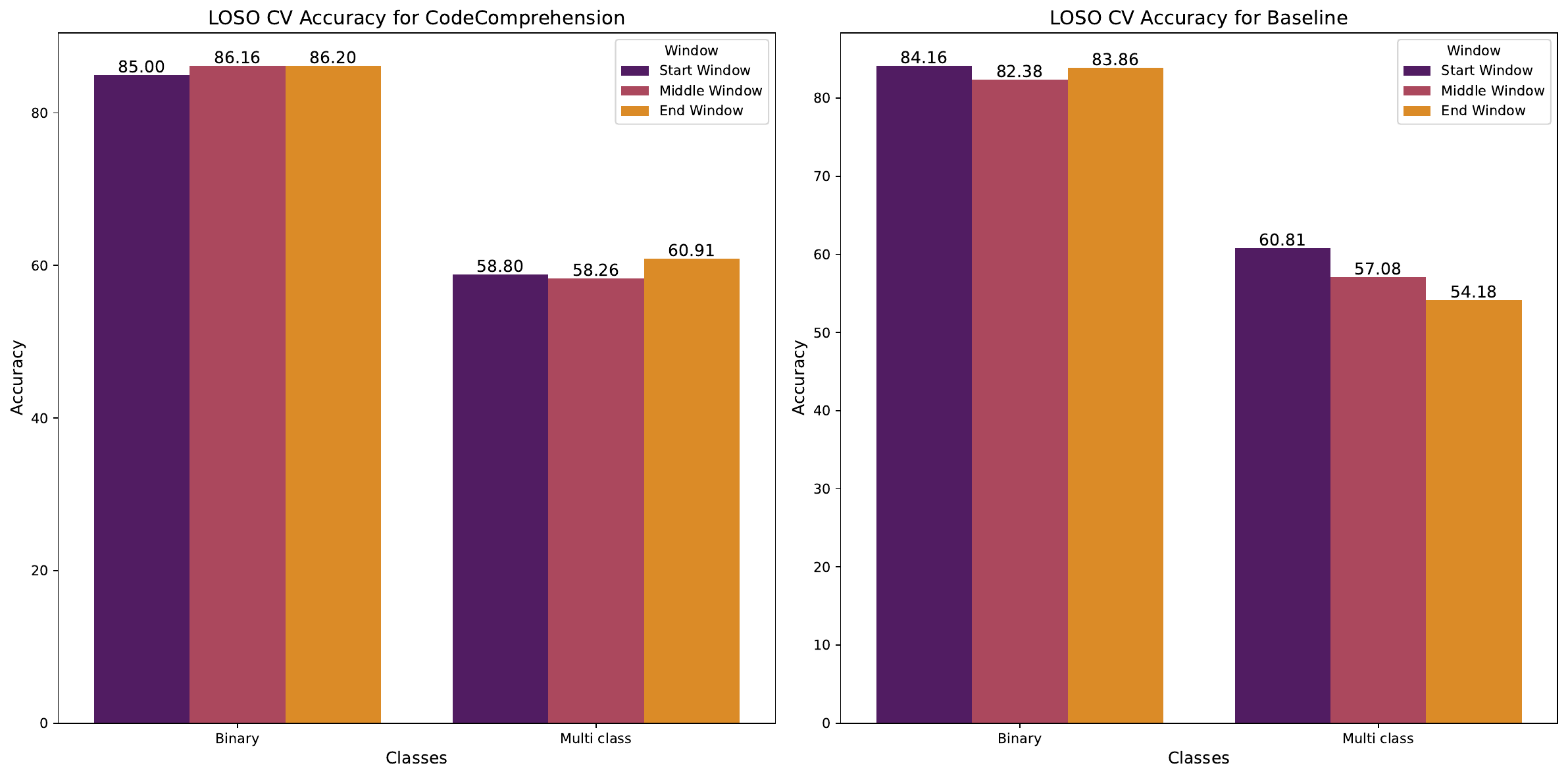}
     \caption{Comparison of different time window positions start, middle and end with leave one subject out evaluation for binary and multi-class classification. Program-comprehension data results are shown on the left, baseline data results on the right side.}    \label{fig:loso_win_results}
 \end{figure*}
 
Figure \ref{fig:loso_win_results} and Figure~\ref{fig:skf_win_results} show the results of this analysis. In the binary classification paradigm, accuracies remained remarkably stable across all windows. For example, in the LOSO baseline condition, the accuracy only marginally fluctuated from 84.16\% (start) to 83.86\% (end). This stability indicates that the neural markers distinguishing the two extreme skill levels novice vs. expert are consistently present throughout the recording.
In contrast, the multi-class classification revealed a clear temporal trend, particularly in the more challenging LOSO validation. In the baseline condition, the classification accuracy showed a notable decline from the start Window (60.81\%) to the end Window (54.18\%). Conversely, for the program-comprehension task, the end window consistently yielded the highest performance in three out of four experimental conditions (e.g., 60.91\% in LOSO multi-class), suggesting that task-specific neural patterns become more pronounced as the cognitive immersion progresses. As participants move past the initial orientation phase of a code snippet, the expertise-specific processing strategies become more distinct, allowing for a more accurate classification of the skill level toward the end of the analyzed epoch.

\section{Discussion}
In this section, we discuss our results in relation to our three research goals.

\subsection{Correlation of Programmer Expertise and EEG Features}

The correlation analysis identifies entropy as the most powerful and consistent predictor of programming expertise across all frequency bands ($\alpha, \beta, \theta, \gamma$). With a Spearman coefficient of $\rho \approx 0.6$. This finding suggests that the fundamental difference between experts and novices lies in the signal complexity of their neural activity. The high positive correlation with entropy indicates that experts exhibit higher signal complexity. In the context of our CSP results, this can be interpreted as a sign of neural efficiency, similar to the findings of \citet{Siegmund:2017:Measuring}. While novices show widespread, perhaps more chaotic or less organized activation, the higher entropy in experts may reflect a more sophisticated and diverse recruitment of localized neural networks. This complexity is not noise, but rather a dense, highly organized information flow within the stable centro-frontal hubs identified in the expert CSP plots.

The analysis highlights $\alpha$ and $\gamma$ attributes (Mean, Variance, StdDev) as stable markers of skill level. Traditionally $\alpha$-Waves are associated with cognitive load and attention. The significant correlation here supports the idea that experts maintain a more controlled state of attention. $\gamma$-waves are often linked to high-level information processing and binding of different information types. Their strong correlation across both baseline and task data confirms that experts possess a different fundamental processing frequency when dealing with code-related logic.

\subsection{Topographic Analysis of Programming Expertise}

The CSP analysis identifies the frontal and centro-frontal regions as the primary topographical markers for program comprehension. While expertise leads to a more focused centro-frontal engagement, the spatial overlap between baseline and task components, particularly in the expert group, suggests that the 30-second rest phase may not represent a completely decoupled physiological state. This may be due to several factors, one of them being cognitive carry-over. A rapid neural transition from complex problem-solving back to a baseline state is challenging, especially for experts whose resting thoughts may remain task-oriented. Another reason might be neural efficiency. Experts may comprehend source code using networks that are already partially active during readiness or resting states. In addition, methodological factors might play a role. The choice of a 4-second epoch from the middle of the rest phase might still capture transitional cognitive activity. 
Future studies shall use longer inter-stimulus intervals to ensure a clear separation of physiological states.

\subsection{Machine-Learning Classification of Programmer Expertise}

The stark contrast between binary and multi-class performance provides significant insight into the nature of programming expertise. While the model highly effectively distinguishes between the extremes (bottom 33\% novices vs. top 33\% experts), the inclusion of the Intermediate group introduces significant classification challenges.
This performance drop is likely explained by the skill score distribution shown in Figure \ref{fig:skillScoreCombined}. The violin plot on the right side reveals a dense concentration of participants around the median with approximately 1.2 correct answers per minute. There is a substantial overlap between those who fall into the intermediate category and those at the boundaries of the expert and novice groups. This suggests that programming expertise exists on a continuous spectrum rather than in discrete, easily separable stages.
Participants in this middle tier may employ a mix of expert-like automated processing and novice-like analytical searching, making them harder to classify based on a universal neural pattern. Future work could benefit from a regression-based approach to expertise rather than a categorical one, given the continuous nature of the observed behavioral data.

The high degree of consistency between task-focused and electrode-focused results is noteworthy. From a neurophysiological perspective, this suggests that the discriminative information is globally distributed across the extracted features rather than being tied to a specific aggregation strategy.
The fact that electrode-focused data did not yield significantly higher accuracy indicates that the spatial information provided by individual channels does not add substantial unique variance over the task-aggregated features in this specific setup. This reinforces the robustness of our classification pipeline, as it implies that the identified expertise markers are stable enough to be captured regardless of whether the analysis emphasizes the temporal task structure or the spatial electrode arrangement. For future applications, this could simplify the data processing pipeline, as simpler aggregation methods may suffice to achieve high classification performance.

The fact that baseline results were nearly as predictive as program comprehension results is particularly interesting. 
One interpretation is that the 30-second rest period was likely insufficient for a full physiological reset. 
The stability of the classification results across the start, middle, and end windows provides critical insights into this issue. 
The high and stable accuracy in the binary paradigm suggests that the differences between novices and experts are so fundamental that they function as stable neural traits. The classifier reaches a performance ceiling where subtle temporal changes, such as the dissipation of task-related activity, do not significantly impair the ability to distinguish these two polar groups.
The multi-class LOSO results act as a stress test for the model. Because the intermediate group shares a significant behavioral and neural overlap with both novices and experts (as seen in the skill score distribution, Figure~\ref{fig:skillScoreCombined}), the classifier relies on more subtle state information. The drop in baseline accuracy toward the end window supports the hypothesis of neural carry-over. Lingering task-related activity in the start window likely aids the differentiation of the ambiguous intermediate group. As this residual activity fades toward the end window, the classification becomes more difficult.
However, the generally high accuracy in the end window of the baseline implies that the neural markers of expertise do not simply fade away within seconds of stopping a task. Instead, these markers may represent a stable, expertise-specific baseline configuration that persists regardless of the immediate temporal distance from the cognitive load, or at least lasts longer the more experienced the programmers are.
It implies that expertise-related differences are not merely task-reactive but are rooted in stable, long-term neural traits or resting-state signatures developed through years of practice.
However, this claim needs further validation in a separate study design with different baseline conditions, involving for example longer baseline periods clearly decoupled from the program-comprehension tasks.

\subsection{Synthesis of Correlation, CSP, and Classification Results}

A crucial observation of this study is that the results for the calculated correlations, common spatial patterns, and classification are nearly identical for both baseline and program-comprehension data. The common patterns and correlations in both conditions explain why our classifier was able to achieve over 84\% accuracy even during rest periods. It reinforces the neural fingerprint hypothesis. Expertise is not just a reactive state triggered by code, but a permanent alteration in how the brain organizes information, visible even in the complexity (entropy) of the resting state. The consistency across task-based and electrode-based arrangements further proves that these features are robust and not artifacts of a specific data-grouping strategy.

The lack of significant correlation in $\theta$ and $\beta$ kurtosis or mean values explains why certain features could be pruned without losing classification power. The ``density of significant correlations'' increasing in electrode-focused data suggests that while the signal is globally present, its spatial distribution, as seen in CSP, provides the most refined evidence of the expertise-specific neural architecture.

The high classification accuracy and the robustness across different aggregation methods, task- vs. electrode-focused, are highly consistent with the spatial patterns identified in the CSP analysis. 
The marginal difference between electrode-focused and task-focused results suggests that the expertise-related neural markers are spatially stable and consistently present throughout the recording. This aligns with the CSP findings for the expert group, which displayed remarkably constant centro-frontal patterns across both baseline and task states. Furthermore, the high classification performance during the baseline period provides empirical support for our observation that baseline and program-comprehension patterns share significant topographical overlaps. 

This leads to the conclusion that programming expertise is characterized by a neural fingerprint, specifically a localized and efficient centro-frontal configuration, that is stable enough to be detected with high precision, regardless of whether the data is aggregated by electrode location or task duration.

\subsection{Limitations and Implications for Application}
Although our study covered a wide range of programmer experience, our conclusions can only be carefully generalized beyond our lab setting. The experiment design required a controlled program-comprehension task rather than a more complex, tool-supported workflow in an industrial software development setting. Finally, our performance-based skill score (correct answers per minute) only considers comprehension efficacy and does not evaluate other practically relevant criteria in software development (e.g., architectural design understanding).

Despite these limitations, our results provide robust empirical evidence that EEG can be used as a complementary perspective to programming expertise. Neural signatures, especially entropy, reveal distinct aspects of cognitive effort and neural efficiency that cannot be inferred from pure behavioral tests. These deeper insights allow opening the black box of programmers' minds~\cite{Peitek:2020}, which paves the way for individualized configuration in research and education. For example, learning programming has been a historically challenging task for students and teachers~\cite{Cheah2020}. EEG-based indicators of high cognitive load during program comprehension could inform adaptive tutoring systems that modulate task difficulty, pacing, or scaffolding for each individual student.

In industrial software development, monitoring of neural signatures could help to detect fatigue, and, for example, prevent introducing bugs before they are even introduced in a software project. Müller and Fritz have already explored an EEG-based monitoring system to detect periodically assess programmers' emotions and progress~\cite{Muller2015}. Our selected EEG features would provide even more meaningful insights into professional workflows.

Another challenge of evaluating expertise in software development is during the hiring process, where common approaches, such as whiteboard interviews, are commonly criticized and poor indicators~\cite{Behroozi2019}. An EEG-based evaluation requires ethical safeguards, but could objectively, accurately, and in an unbiased way inform about a candidate's expertise, without immediate performance pressure common in other settings.

\section{Conclusion}

This study investigated the neural signatures of programming expertise by analyzing EEG data from participants of varying skill levels during baseline and program-comprehension tasks. By combining spatial, statistical, and machine learning analyses, we provided a comprehensive overview of how professional experience reshapes the programmer's brain. Our topographical analysis (CSP) revealed a clear progression in neural engagement: While novices rely on a distributed network involving frontal and parietal regions—likely reflecting high visual search effort and syntax decoding—experts exhibit a highly localized and stable centro-frontal focus. This suggests that expertise is characterized by increased neural efficiency and the automation of cognitive processes. The correlation analysis identified entropy across all frequency bands as the most robust predictor of skill level. The higher signal complexity in experts, particularly in the $\alpha$ and $\gamma$ bands, serves as a reliable marker for sophisticated information processing. These findings were further validated by our classification results, which achieved a high binary accuracy of up to 92\%. A pivotal finding of this work is the persistence of expertise-related patterns during the resting state. The consistently high classification performance in the baseline data, coupled with the temporal analysis of various windows, indicates that programming expertise is not merely a transient task-related state but a stable neural trait. However, the observed decline in multiclass accuracy toward the end of the baseline suggests that a 30-second rest period is insufficient for a complete physiological reset, as task-related activity lingers. In summary, this research demonstrates that the programmer's skill level can be accurately decoded from EEG signals using established machine learning methods like Random Forest. Our results emphasize that future studies should account for the temporal dynamics of the baseline and consider expertise as a continuous spectrum rather than in discrete, easily separable stages. These insights contribute to a deeper understanding of cognitive load and expertise development in software engineering, potentially informing future neuro-adaptive educational tools.

\subsection{Acknowledgments}
\noindent 
We thank all participants of the study of Peitek et al.~\cite{Peitek:2022:Correlates}.

\subsection{Funding}
\noindent This work has been supported by several national grants. 

\subsection{Data availability}
\noindent The original EEG data of Peitek et al.~\cite{Peitek:2022:Correlates} is available: \url{https://osf.io/4hjbd/}. Our analysis code will be made publicly available upon publication.

\bibliographystyle{IEEEtranN}
\bibliography{references}\emph{}

\end{document}